\documentclass[amsmath,twocolumn,superscriptaddress,prl,aps]{revtex4}
\usepackage{amsthm,amsfonts,graphicx,verbatim}
\newcommand{\be}{\begin{equation}}
\newcommand{\ee}{\end{equation}}
\newcommand{\bea}{\begin{eqnarray}}
\newcommand{\eea}{\end{eqnarray}}

\newcommand{\la}{\langle}
\newcommand{\ra}{\rangle}
\newcommand{\lb}{\left[}
\newcommand{\rb}{\right]}
\newcommand{\lp}{\left(}
\newcommand{\rp}{\right)}

\newcommand{\Tr}{{\rm Tr}\,}
\renewcommand{\phi}{\varphi}
\renewcommand{\epsilon}{\varepsilon}
\renewcommand{\vec}[1]{{\bf #1}}
\renewcommand{\Im}{{\rm Im}\,}

\begin{document}

\title{
% Bragg Scattering,
Peierls-type Instability and Tunable Band Gap in Functionalized Graphene}
\author{D. A. Abanin}
\affiliation{Princeton Center for Theoretical Science and Department of Physics, Princeton University, Princeton NJ 08544, USA}
\author{A. V. Shytov}
\affiliation{Department of Physics, University of Exeter, Exeter, EX4 4QL, U.K.}
\author{L. S. Levitov}
\affiliation{Department of Physics, Massachusetts Institute of Technology, Cambridge MA 02139, USA}

%\date{\today}

\begin{abstract}

Functionalizing graphene was recently shown to have a dramatic effect on the electronic properties of this material. Here we investigate spatial ordering of adatoms driven by the RKKY-type interactions.
In the ordered state, which arises via a Peierls-instability-type mechanism, the adatoms reside mainly on one of the two graphene sublattices.
Bragg scattering of electron waves induced by sublattice symmetry breaking results in a band gap opening, whereby Dirac fermions acquire a finite mass. The band gap is found to be immune to the adatoms' positional disorder, with
only an exponentially small number of localized states residing in the gap. The gapped state is stabilized in a wide range of electron doping.
Our findings show that controlled adsorption of adatoms or molecules provides a route to engineering a tunable band gap in graphene.

\end{abstract}

\maketitle

The unique electronic properties of graphene, a one-atom-thin carbon sheet with a tunable electron density~\cite{Novoselov04} and high carrier mobility~\cite{Chen08,Bolotin08}, make it an attractive material for applications in nano-electronics~\cite{Novoselov07}. However, because of the gapless semi-metallic character of graphene band structure, the future of graphene electronics depends on developing methods to engineer a band gap in this material. The gapless character of electron dispersion in pristine graphene is protected by the high symmetry of its lattice, in which two carbon sites in the unit cell are equivalent. The simplest kind of gap-opening perturbation which lifts this symmetry can be described by unequal potentials $u_A$ and $u_B$ on the $A$ and $B$ sites \cite{CastroNeto09}, leading to a finite mass of Dirac quasiparticles near points $K$ and $K'$  of the Brillouin zone. The quasiparticle spectrum, described by the Hamiltonian
\be\label{eq:hamiltonian}
H_{K(K')}=\left[\begin{array}{cc}
         u_A  &  v_0 (p_1\!\pm \!ip_2)  \\
         v_0 (p_1\!\mp \! ip_2)&  u_B
      \end{array}
\right]
% , \quad p_{\pm}=p_x\pm i p_y
,\quad
v_0\approx 10^6 {\rm m/s}
,
\ee
features a band gap of size $\Delta=|u_A-u_B|$, which opens due to Bragg scattering  of electron waves on the periodic sublattice potential.

A gap opening via such a mechanism could occur in epitaxial graphene, grown or placed on a lattice-matched substrate~\cite{deHeer04,Lanzara07,Giovannetti07}. Yet, while the approach involving lattice-matched substrates is simple and direct, combining it with transport measurements proved challenging (see also Ref.~\cite{Rotenberg08}). A gap opening due to
% potential difference between sites $A$ and $B$
sublattice asymmetry is more readily achievable in bilayer graphene, where the sites $A$ and $B$ reside on different layers.  In bilayer graphene, the $A/B$ symmetry can be lifted by asymmetric chemical doping~\cite{Ohta06,Castro07} or electrical gating~\cite{Oostinga08}, leading to a gap opening.

\begin{figure}
\includegraphics[width=3.4in]{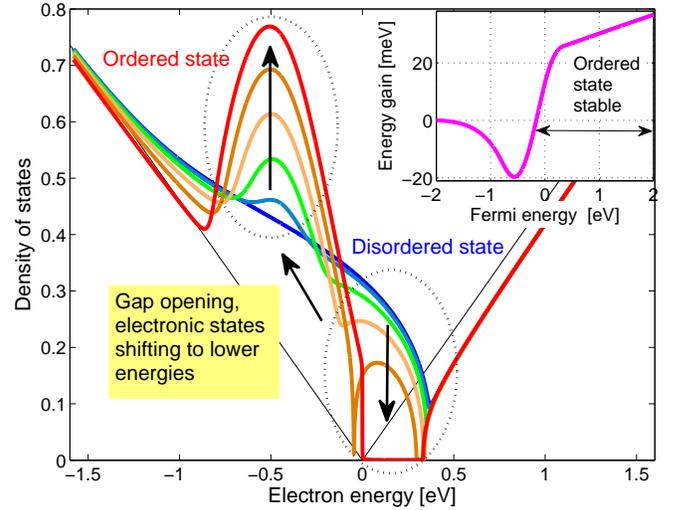}
\vspace{-4mm}
\caption[]{Peierls-type instability resulting from adatom ordering over sublattices $A$ and $B$. A gap in the density of electronic states opens up due to Bragg scattering on the $A/B$ modulation when the occupation probabilities are unequal,  $n_A\ne n_B$. The states in the gap \emph{move down in energy} into a peak positioned at the energy of a single adatom resonance. Inset: the ordered state is stabilized in a wide range of carrier densities, for which the energy gain per adatom is positive. For details of calculation see discussion following Eq.(\ref{eq:T}). Different curves correspond to the occupancy fraction $n_A/(n_A+n_B)=0,0.1,...,0.5$ with $(n_A+n_B)/2=0.04$, $U=6t_0$, $W=3t_0$.
%for different concentration of adatoms on the two sublattices, calculated within the self-consistent Born approximation, Eq.(\ref{eq:DOS}). In the sublattice-imbalanced case, $n_A\neq n_B$, a band gap opens up, while in the sublattice-balanced case, $n_A=n_B$, the DOS is finite at all energies. Parameters used: $n=(n_A+n_B)/2=0.04$, $U=6t_0$, $W=3t_0$.
}
\vspace{-4mm}
\label{fig0}
\end{figure}

Another promising method for gap engineering relies on spatial confinement,
involving patterning graphene into narrow ribbons~\cite{Louie06,Han07}, or quantum dots~\cite{Ponomarenko08}. The gap obtained by such a method can be tuned by varying spatial width of graphene ribbons or dots. However, the approaches relying on spatial confinement are prone to disorder, because of scattering of electron waves on rough edges of patterned graphene. Localized states, appearing inside the band gap, transforms
it into a ``transport gap'' \cite{Adam08}. In contrast, the gap opened due to lifting the $A/B$ symmetry can be expected to be more robust in the presence of disorder, as long as the mean free path is large compared to the $A/B$ modulation period.

An elegant approach to modify electronic properties of graphene, demonstrated recently \cite{Elias09,Rotenberg09}, is based on the well established technique of chemical functionalization, in which groups such as H, OH, or F bind covalently to carbon atoms, transforming the trigonal $sp^2$ orbital to the tetragonal $sp^3$ orbital. Such transformation drastically alters local electronic properties. Theory predicts that, at $100\%$ coverage by H adatoms, graphene turns into a wide-gap semiconductor, called graphane \cite{Sofo07}. The experiments~\cite{Elias09,Rotenberg09}, however, are done at low coverage, typically of about few percent.
Can a state with a band gap be realized in the low-coverage regime?

Electronic properties at low adatom coverage are dominated by resonant scattering of electron waves on the adatoms~\cite{Pereira06,Pereira08,Wehling08}. Pairwise RKKY-type interactions between adatoms were analyzed in Ref.~\cite{Shytov09}, where the interaction sign was found to depend on whether the interacting atoms occupy the same sublattice or different sublattices. Such sublattice dependence suggests that the RKKY interactions can drive ordering of the adatoms in which sublattices $A$ and $B$ become unequally populated.
% In the resonant scattering model, which is appropriate for H adatoms, the electron-mediated interaction is of a long-range character, $U(r)\propto 1/r$, which makes its effect significant even at low coverage.

Here we propose a mechanism for spontaneous ordering, illustrated in Fig.\ref{fig0}, which is analogous to that of Peierls instability. The adatom ordering over sublattices $A$ and $B$ leads to a gap opening due to electron waves Bragg scattering on the $A/B$ modulation, resulting in electronic states in the gap shifting up and down in energy. Crucially, these shifts are asymmetric, with states \emph{shifting predominatly down in energy} to a peak centered at the energy of a single atom resonance, $\epsilon=\epsilon_0<0$.  The system gains energy as a result of such level shifts for electron dopings in the range indicated in Fig.\ref{fig0} inset, corresponding to positive chemical potential values. For such dopings, the gapped state with unequal sublattice population is stabilized. The gap value is determined by the scattering properties of adatoms and their concentration, and is therefore \emph{tunable}.

Because of the resonant character of electron scattering, the electron-mediated interactions fall off slowly with distance at adatom separations $r\lesssim \ell_0=\hbar v_0/|\epsilon_0|$, as $U(r)\sim 1/r$, and more rapidly at larger distances~\cite{Shytov09}. Hence for not too low adatom coverage, $n\gtrsim n_*=(a/\ell_0)^2$, where $a=0.142\, {\rm nm}$ is the lattice constant, the adatom ordering cannot be analyzed using a pairwise interaction model. Here we present a theory which fully accounts for the non-pairwise, collective nature of electron-mediated interactions in functionalized graphene.

Our approach applies to different atoms and chemical groups used to functionalize graphene.
{\it Ab initio} study~\cite{Wehling08} predicts the resonance energy values which span a wide range:
%that possible values of $\epsilon_0$ span a broad range:
$-\epsilon_0=0.03,\, 0.11,\, 0.70,\,0.67\,{\rm eV}$ for H, CH$_3$, OH, and F, respectively. This corresponds to the characteristic values $n_*\approx 10^{-4},\,10^{-3},\,0.05,\,0.05$. In the limit of very small coverage, $n\ll n_*$, a pairwise interaction model can be used to describe ordering\,\cite{Cheianov09,Cheianov10}, whereas for larger coverage values a self-consistent treatment presented below must be employed.

\begin{figure}
\includegraphics[width=3.4in]{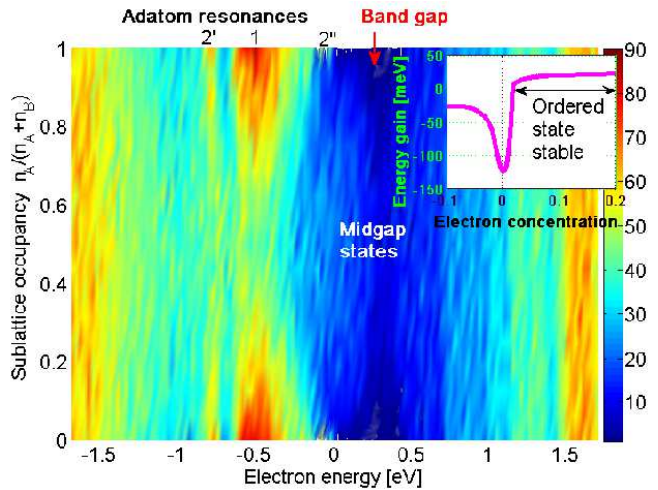}
\vspace{-4mm}
\caption[]{The density of electronic states as a function of energy for different sublattice occupancy ratios, obtained by numerical diagonalization of the Hamiltonian (\ref{eq:hamiltonian1}), averaged over 160 realizations of disorder. The band gap, which opens at $n_A\ll n_B$ and $n_B\ll n_A$, is immune to disorder: no electronic states are found inside the gap. Sublattice ordering results in the energy gain at positive dopings (inset). Parameters used: system size $62\times 34$, $U=6t_0$, $n=(n_A+n_B)/2=0.034$.}
\vspace{-4mm}
\label{fig_simulation}
\end{figure}

The Peierls-type scenario for ordering described above
% presented in Fig.\ref{fig0}
can be tested by direct numerical diagonalization of the nearest neighbor tight binding Hamiltonian (see Fig.\ref{fig_simulation}):
\be\label{eq:hamiltonian1}
H=\sum_{|\vec x-\vec x'|=1} t_0 (\psi^{\dagger}_{\bf x}\psi_{\bf x'}+{\rm h.c.})+\sum_{\bf x} u({\bf x})\psi^{\dagger}_{\bf x} \psi_{\bf x}
,
\ee
with $t_0\approx 3.1\, {\rm eV}$, and potential $u(\vec x)=\sum_i U\delta(\vec x-\vec x_i)$ taking value $U$ on the sites occupied by adatoms. Large $U\gg t_0$ was used to model the effect of the $sp^2$ to  $sp^3$ transformation, which inhibits the conduction electrons from occupying the adatom sites, effectively turning these sites into vacancies. In the simulation shown in Fig.\ref{fig_simulation} we used $U=6t_0$, which gives the resonance energy positioned at $\epsilon_0\approx -0.4\,{\rm eV}$.
% The density of states (DOS), obtained for different occupancy values by diagonalizing the Hamiltonian (\ref{eq:hamiltonian1}), is shown in Fig.\ref{fig_simulation}.

The behavior of the DOS, obtained for different occupancy values by diagonalizing the Hamiltonian (\ref{eq:hamiltonian1}),
agrees well with the results obtained by an analytic method (see Fig.\ref{fig0}). The peak at $\epsilon= \epsilon_0$, which is a signature of resonant scattering on individual adatoms~\cite{Pereira06,Wehling08},
%, which would be a zero-energy mode for $U=\infty$~\cite{Pereira06,Wehling08},
is present for all occupancies, but is more pronounced for the sublattice-ordered state, $n_A\gg n_B$ or $n_B\gg n_A$. The resonances marked $2'$ and $2''$ correspond to the single-particle states formed near two neighboring adatoms. The DOS remains finite at all energies for $n_A\approx n_B$. In contrast, the DOS vanishes in the interval $0<\epsilon\lesssim 0.4 eV$ for the sublattice-ordered state, which corresponds to the band gap opening.

To estimate the energy gain due to ordering, we evaluate the energy of the system as a function of the adatom occupancy fraction and electron concentration $N/N_{s}-1/2$ ($N$ is the number of electrons, $N_s$ is the number of sites).
% \mpar{LL: spin degeneracy not included?}
The results for $\delta E$, the energy gain per adatom, shown in Fig.\ref{fig_simulation} inset, agree with our Peierls-type argument: the sublattice-ordered state is stabilized at positive doping. The ordering temperature, estimated as $T_c=2\delta E$ (see Appendix), takes values 
in the hundreds of Kelvin for parameters used in Fig.~\ref{fig_simulation}.

\begin{figure}
\includegraphics[width=2.2in]{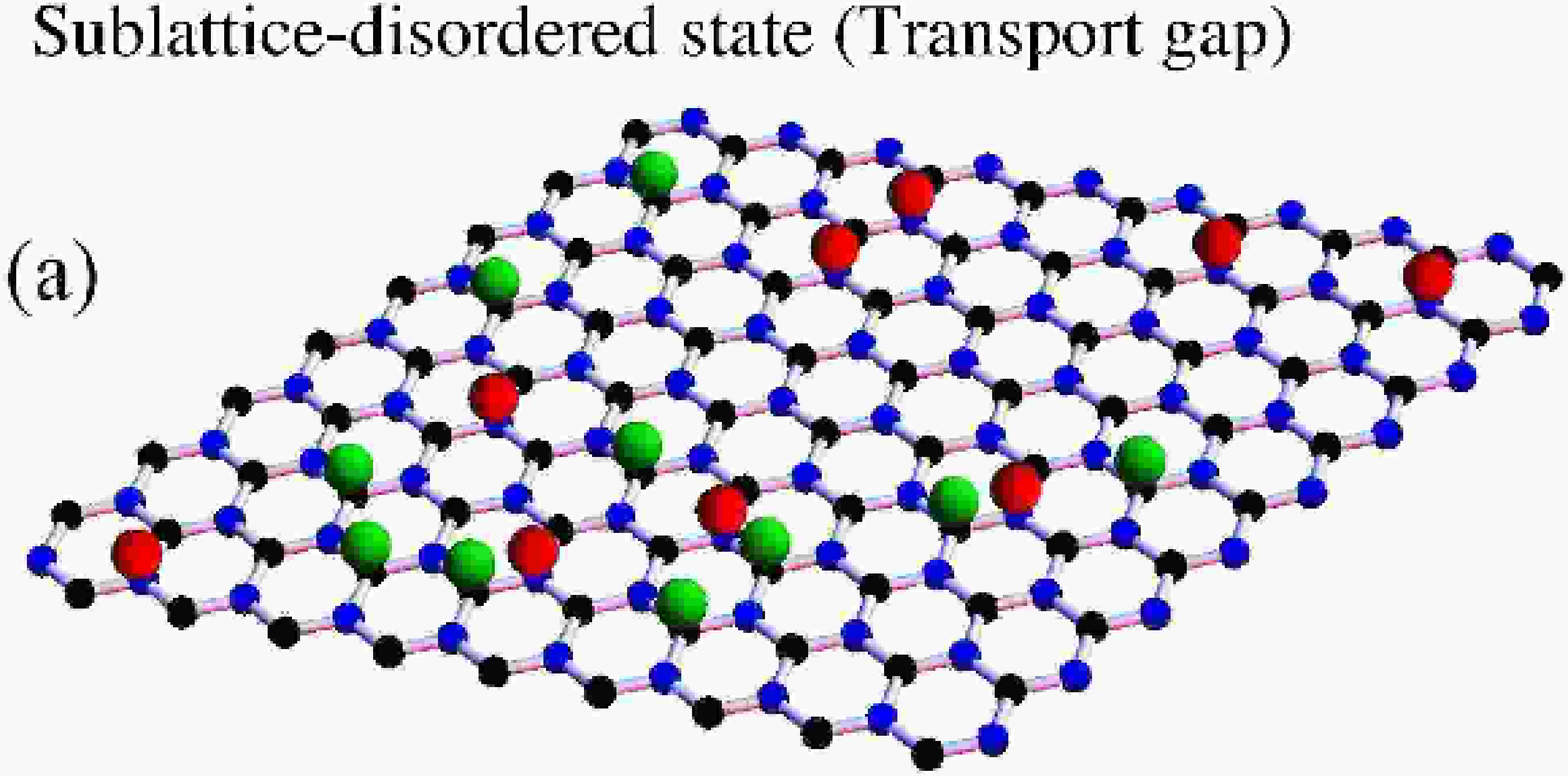}
\includegraphics[width=0.9in]{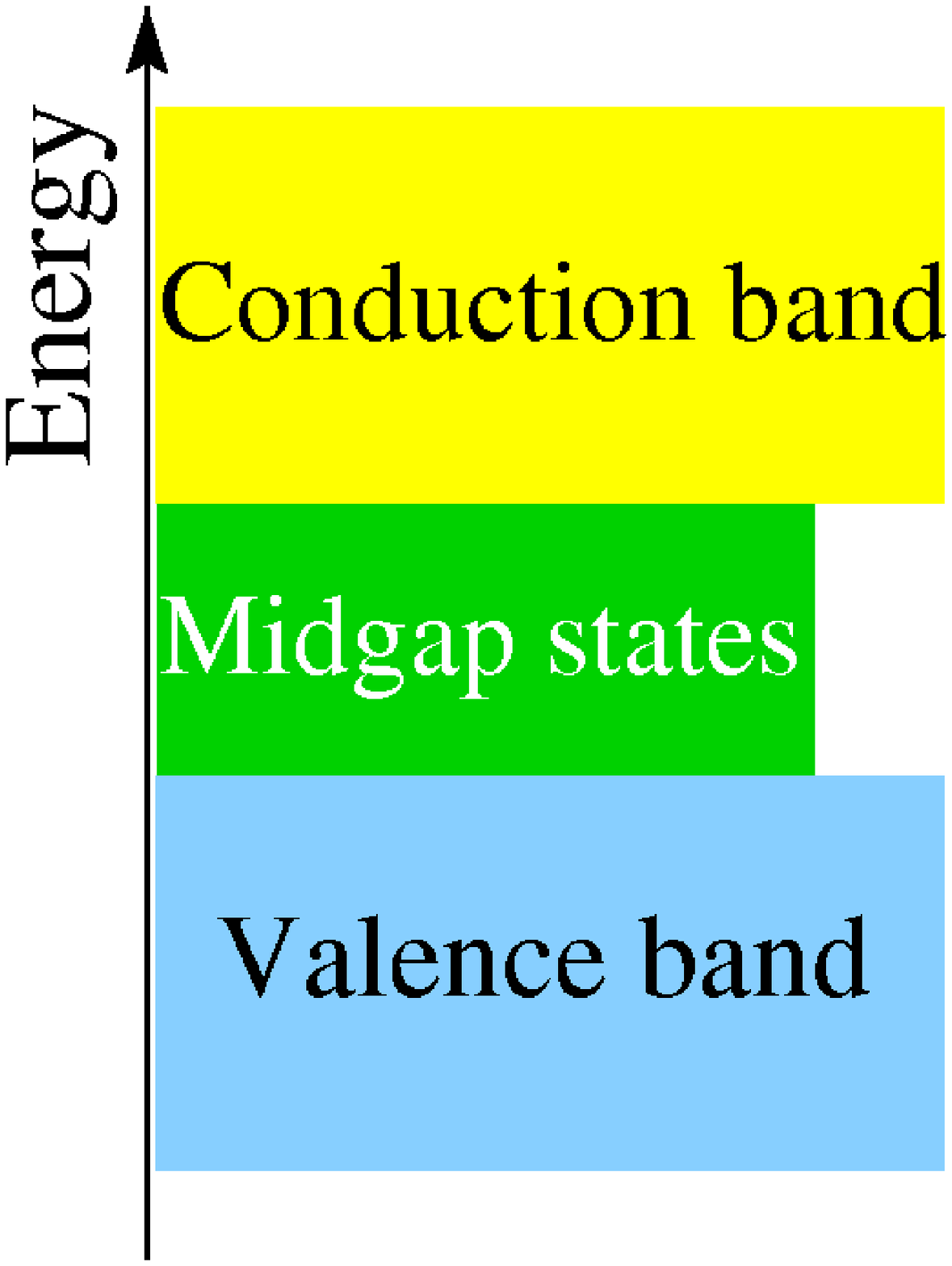}

\includegraphics[width=2.2in]{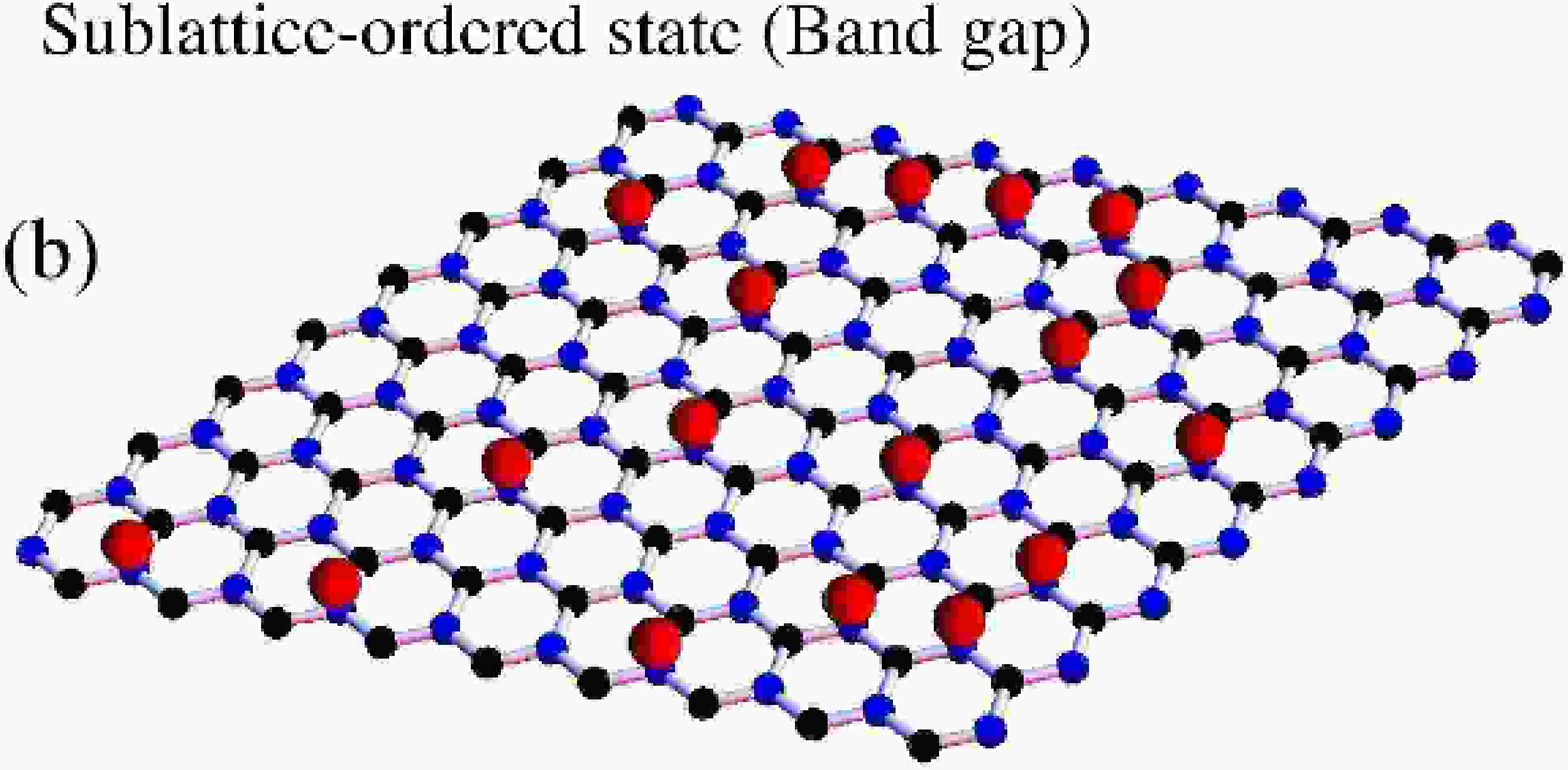}
\includegraphics[width=0.9in]{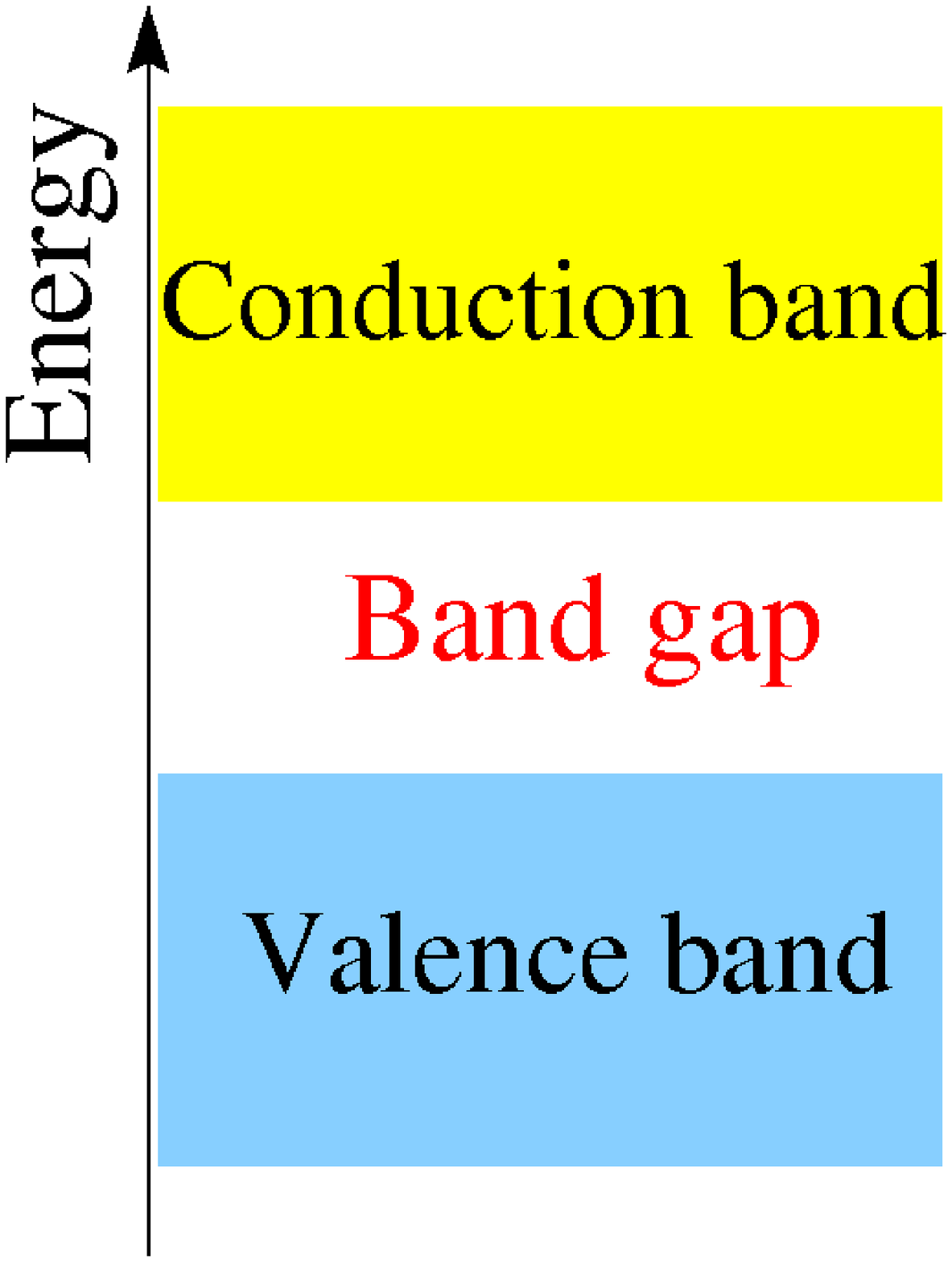}
\vspace{-4mm}
\caption[]{Two phases of adatoms on graphene: disordered (a), where adatoms are randomly distributed over two sublattices of the hexagonal lattice, $A$ and $B$, and ordered (b), where adatoms preferentially occupy one of the sublattices. The adatoms residing on sublattice $A(B)$ are shown in red (green). Schematics of the energy spectrum in the two phases are also shown. In the ordered phase, the sublattice symmetry breaking leads to a band gap.}
\vspace{-4mm}
\label{fig_lattice}
\end{figure}

Remarkably, in our simulation the energy gap is found to be immune to the effects of disorder. As Fig.\ref{fig_simulation} indicates,  no states inside the gap are found when the adatoms randomly populate one sublattice, $A$ or $B$ (160 disorder realizations were analyzed). Such partial ordering, illustrated in Fig.\ref{fig_lattice}, is sufficient to completely expel the midgap states, and open a band gap.
This surprising result seems to be in an apparent contradiction with the intuition based on the physics of localization of electronic states in disordered systems.

To understand this behavior, we shall start with a simple case of a weak adatom potential $U\ll t_0$, and then generalize to the case of strong potential. At a weak potential, within the mean-field approximation, we have $u_A=n_AU$ and $u_B=n_BU$ in the Hamiltonian (\ref{eq:hamiltonian}). For the gap to survive in the presence of disorder, its value should exceed the disorder broadening, $\Delta=|u_A-u_B| \gg \hbar/\tau_{\epsilon\sim\Delta}$. Using the Born approximation for the scattering rate, we find
\be\label{eq:scattering_time}
\frac{\hbar}{\tau(\epsilon)}=\frac{\pi}{2 n_0} \rho(\epsilon) |U|^2 (n_A+n_B)
,\quad n_0=\frac2{3\sqrt{3}a^2}
% \tilde{U}=\frac{U}{n_0}
,
\ee
where
% $n_{A(B)}^*=n_{A(B)}n_0$, and $n_0=4/(3^{3/2}a^2)$
$2n_0$ is the density of carbon atoms in the graphene lattice.
%, and $a=0.142\, {\rm nm}$ is the lattice constant.
Taking $\rho(\epsilon)$ to be that of pure graphene for one spin projection, $\rho(\epsilon)=\frac{|\epsilon|}{\pi \hbar^2 v_0^2}$, and using the relation $v_0=3t_0a/2\hbar$, we rewrite the condition
$\Delta \gg \hbar/\tau_{\epsilon \sim \Delta}$ as
\be\label{eq:condition}
(n_A+n_B)U^2\ll t_0^2.
\ee
This condition is always satisfied for weak adatom potential $U\ll t_0$. 

This argument can also be applied, with a slight modification, to the case of strong 
adatom potential, $U\gg t_0$. This can be done by replacing $U$ in Eq.(\ref{eq:scattering_time}) by a suitably defined $T$-matrix, see Eq.(\ref{eq:T}). This again yields the condition (\ref{eq:condition}), which is satisfied at low enough adatom coverage. Furthermore, as numerical results presented in Fig.~\ref{fig_simulation} show, the gap persists even at higher adatom concentration, $(n_A+n_B)U^2\sim t_0^2$.

Of course, strictly speaking, the DOS inside the gap must be nonzero. However, since the states deep in the gap can arise only due to relatively large fluctuations of disorder, their contribution to the DOS is exponentially small. To analyze this quantitatively, we shall focus on the simplest case of a weak potential, $U\ll t_0$, assuming that all adatoms reside on the $A$ sublattice. We model the effect of disorder by the Hamiltonian (\ref{eq:hamiltonian}) with fluctuating gap $\Delta(\vec r)=u_A=n_A(\vec r) U$, and zero $u_B$.

This problem can be mapped on the well studied problem of the DOS
below the band edge in a disordered semiconductor~\cite{Halperin66,Zittartz66}. Starting with the equations $\epsilon \psi_A=\Delta(\vec r)\psi_A+v_0p_+\psi_B$, $\epsilon \psi_B=v_0p_-\psi_A$, and eliminating the component $\psi_B$, we obtain an eigenvalue equation
\be\label{eq:shroedinger}
\lp \epsilon  \Delta(\vec r)-v_0^2 \nabla^2\rp \psi_A(\vec r)=\epsilon ^2 \psi_A(\vec r).
\ee
For $\epsilon$ near the upper band edge, given by the disorder-averaged potential, $\bar \Delta=n_A U$, we expand in $\delta \epsilon =\epsilon -\bar \Delta$ to bring the eigenvalue equation to the form of the Schroedinger equation for a massive non-relativistic particle:
\be\label{eq:ham_random}
\delta \epsilon \psi_A(\vec r)=\lp -\frac{\nabla^2}{2m}+\delta u(\vec r)\rp \psi_A(\vec r)
,\quad
m=\frac{\bar\Delta}{2v_0^2}
,
\ee
where $\delta u(\vec r)=\delta n_A(\vec r) U$ is the fluctuating part of the gap $\Delta(\vec r)$. We treat the long wavelength fluctuations of $\Delta(\vec r)$ as Gaussian with the two-point correlation function $\la \delta u(\vec r)\delta u(\vec r')  \ra= U^2\la \delta n_A(\vec r)\delta n_A(\vec r')\ra=\lambda\delta(\vec r-\vec r')$, where $\lambda=U^2 n_A/n_0$, as appropriate for delta-correlated adatom positions.
In this case the DOS for the problem (\ref{eq:ham_random}) decays exponentially away from the band edge~\cite{Zittartz66},
\be\label{eq:dos_midgap}
\rho(0<\epsilon \lesssim \bar u_A)\propto \exp(-c |\delta\epsilon| /\xi)
,\quad
\xi=\lambda m
\ee
%
%\mpar{Corrected formula for $\xi$}
with $c$ a constant of order one. Estimating the energy scale $\xi$, we see that, at low coverage $n_A\ll1$, it is much smaller than the gap width: $\xi=\lambda m=\frac{U^2n_A \Delta}{2 n_0 v_0^2}\sim \frac{\Delta ^2}{t_0^2}U \ll \Delta$. Therefore we conclude that the DOS is exponentially small within the gap. This is consistent with the results of our simulation, Fig.\ref{fig_simulation}, in which some smearing of the DOS was observed at the gap edge; however no states were found deep inside the gap for all of the 160 disorder realizations which we analyzed.

An analytic approach to analyze transport properties, such as disorder scattering and conductivity, can be developed using a self-consistent T-matrix approximation (SCTA) (see Appendix).
In the model (\ref{eq:hamiltonian1}), the $T$-matrix of an individual adatom has a resonant form~\cite{Pepin01},
\be\label{eq:T}
T_0(\epsilon)=\frac{\pi v_0^2}{\epsilon \ln (iW/\epsilon)+\delta}
,\quad \delta=\frac{\pi v_0^2}{\tilde U}
,\quad \tilde U=U/n_0
,
\ee
where $W\approx 3t_0$ is the bandwidth, and the parameter $\delta$ determines the energy of the resonance, $\epsilon_0 \ln W/|\epsilon_0|=-\delta$. We note that the description of adatoms by an on-site potential is equivalent to the model which describes adatom states in terms of a localized level hybridized with the graphene continuum \cite{Wehling08}, since the form of the $T$-matrix at low energies is the same in both approaches.

We calculate disorder-averaged Greens functions using the self-consistent
approach (see Appendix). The Greens functions are then used to extract the density of states,  $\rho(\epsilon)=-\frac{1}{\pi}\Im \Tr G(\epsilon+i0)$. The resulting energy dependence of the DOS is shown in Fig.\ref{fig0}. The main features, such as the gap opening for imbalanced $A/B$ occupancies, accompanied by the shift of electronic states into the resonance peak of a single adatom, are in agreement with our numerical results.
The DOS is then used to evaluate the total energy as a function of the chemical potential, $E(\mu)=\int_{-\infty}^{\mu} \epsilon \rho(\epsilon)d\epsilon$.
%{\bf .. and end here}
The change in the electronic energy due to sublattice ordering, $\delta E=E_{n_A=n_B}-E_{n_A\ll n_B}$, depends on $\mu$ as shown in Fig.\ref{fig0} inset. The sublattice-ordered state is stabilized for positive dopings $\mu>0$.

The gap opening and its character reveals itself in transport measurements, since the temperature dependence of conductivity is activation-like in systems with an intrinsic band gap, $\sigma \propto e^{-\Delta/k_{\rm B}T}$, but has a variable-range hopping behavior of the Mott or Efros-Shklovskii form for systems with a transport gap. In addition, sublattice ordering suppresses scattering, which leads to {\it an increase in conductivity} for electron doping above or below the gap~\cite{AbaninLevitov10}. Observation of such an increase can serve as a hallmark of adatom ordering.

Conductivity of the system, evaluated within the SCTA approach, is given by~\cite{AbaninLevitov10}, 
\be
\sigma(\mu)\approx \frac{2e^2}{\pi^3 \hbar^3}\frac{\mu^2\ln^2 W/|\mu |}{v_0^2(\tilde n_A +\tilde n_B)}
+\frac{e^2}{\pi^2\hbar}\frac{(n_A -n_B)^2}{(n_A +n_B)^2}\ln \frac{W}{|\mu |}
,
\ee
which is valid far outside the gap region, $|\mu|\gg \Delta,|\epsilon_0|$. The quadratic dependence on $\mu$ in the leading term is characteristic for resonant scattering~\cite{Ostrovsky06}. The subleading term, proportional to $n_An_B/(n_A+n_B)^2$, is negative. Hence, scattering is indeed suppressed and  {\it conductivity enhanced} due to sublattice ordering.

An experimental approach to realizing the sublattice ordered state depends on the lateral mobility of adatoms. If the adatoms remain mobile below the ordering temperature $T_c=2\delta E$ (see Appendix), the ordering will occur via a conventional Ising-type phase transition. In this case, a rapid cooldown following ordering may be needed to prevent adatom clustering~\cite{Shytov09}. However, if the mobility is quenched at temperatures $T>T_c$, the system must be annealed at $T\lesssim T_c$ to achieve ordering and gap opening. Since only a small part of adatom's entropy needs to be removed for sublattice ordering (see Fig.\ref{fig_lattice}), it should take only a few hops by each adatom to transition into the gapped state.
% \mpar{LL: Ising domains}

In summary,
the interaction between adatoms in functionalized graphene can drive sublattice ordering via a Peierls-type transition. The band gap, opened by Bragg scattering of electron waves on the sublattice modulation, is immune to positional disorder of adatoms, with the density of localized states inside the gap being exponentially small. The gapped state is shown to be stable in a wide range of electron doping. 

We thank A. K. Geim, K. S. Novoselov, and E. Rotenberg for useful discussions. This work was supported by Office of Naval Research Grant No. N00014-09-1-0724.

%% {bibliography}

\section{Appendix}

\subsection{Self-consistent T-matrix approximation}

Here we discuss the self-consistent T-matrix approximation (SCTA).
%We evaluate the DOS and the total energy using Greens functions calculated in SCTA.
We use the Greens function expressed through disorder-averaged self-energy,
% \mpar{LL: changed sign of $t_{\vec p}$}
%
\be\label{eq:Gmatrix}
G(\epsilon,\vec p)=\left[\begin{array}{cc}
         \epsilon_A  & -t_{\vec p}  \\
         -t_{\vec p}^*&  \epsilon_B      \end{array}
\right]^{-1}
, \quad
\epsilon_{A(B)}=\epsilon-\Sigma_{A(B)}(\epsilon)
,
\ee
with $t_{\vec k}=t_0(1+e^{-i{\vec k \vec e_1}}+e^{-i{\vec k\vec e_2}})$. An infinitesimal imaginary part $\pm i0$ should be added to $\epsilon$ to obtain the retarded and advanced Greens functions.

The self-energy is approximated by the average value of the $T$-matrix, separately for the $A$ and $B$-type adatoms,
\be\label{eq:self-energy}
% \Sigma_{A(B)}(\epsilon)=n_{A(B)}^*\la T(\epsilon)\ra _{A(B)}
\Sigma_{A}(\epsilon)=\tilde n_{A}\la T_{A}(\epsilon)\ra
,\quad
\Sigma_{B}(\epsilon)=\tilde n_{B}\la T_{B}(\epsilon)\ra
% \Sigma_{A(B)}(\epsilon)=n_{A(B)}^*T(\epsilon),
,
\ee
with $\tilde n_{A(B)}=n_{A(B)}n_0$ the adatom densities. Here the quantities $T_{A(B)}$, written as a $2\times 2$ matrix, are given by
\be\label{eq:Tmatrix}
% T(\epsilon)=
\left[\begin{array}{cc}
         T_A  & 0  \\
         0&  T_B      \end{array}
\right]=\frac{\tilde{U}}{1-\tilde{U} g}, \,\, g=\int \frac{d^2p}{(2\pi)^2} G(\epsilon,\vec p).
\ee
The integral of the Greens function over the Brillouin zone is dominated by the regions near $K$ and $K'$, giving
%
% In the above equation, the integration is performed near one of the corners $K,K'$ of the Brillouin zone, and the angular integration makes the matrix structure of $g$ trivial. As a next step, we express $g$ in terms of $\epsilon_{A}, \epsilon_B$~\cite{Shytov09},
%
\be\label{eq:g}
g=
-\frac{\ln (-W^2/\epsilon_A \epsilon_B)}{2\pi v_0^2}\left[\begin{array}{cc}
       {\epsilon_B} & 0  \\
         0 &  {\epsilon_A}
      \end{array}
\right]
,\quad W\approx 3t_0
.
\ee
valid for $|\epsilon_{A(B)}|\ll W$. Combining this result with Eq.(\ref{eq:self-energy}), we obtain two coupled equations for $\epsilon_A$, $\epsilon_B$:
\be\label{eq:eAeB}
\epsilon_{A}=\epsilon-\frac{n_{A} U}{1+\gamma \epsilon_{B} }, \,\,\epsilon_{B}=\epsilon-\frac{n_{B} U}{1+\gamma \epsilon_{A} },
\ee
where we used the relation $U=\tilde{U}n_0$, and defined
\be\label{eq:gamma}
\gamma=\frac{1}{\sqrt 3\pi} \frac{ U}{t_0^2} \ln\left( - \frac{W^2}{\epsilon_A \epsilon_B}\right).
\ee
To obtain retarded (advanced) Greens function, one should choose the branch of the logarithm that is analytic in the upper (lower) half-plane.
Solving numerically for $\epsilon_A$, $\epsilon_B$ as a function of $\epsilon$, we find the Greens function (\ref{eq:Gmatrix}) and use it to calculate the density of states,
\be\label{eq:DOS}
\rho(\epsilon)=\frac{1}{\pi}\Im \Tr G(\epsilon+i0).
%\frac{1}{2\pi^2 v_0^2} {\rm Im  }\left[ -i(\epsilon_A+\epsilon_B) \ln\frac{W^2}{\epsilon_A \epsilon_B} \right]  %|_{i\epsilon\to \epsilon+i0}.
\ee
{This approach was used to produce the curves shown in Fig.1 of the main text.}

\subsection{An estimate of the ordering temperature}

In this section, we consider adatom ordering on the sublattices $A$ and $B$, and estimate the ordering temperature. Since the ordering involves breaking of the sublattice symmetry, it can be described as a second-order phase transition.
%LL Finally, we estimate the ordering temperature $T_c$. Adatom ordering, accompanied by the breaking of the sublattice symmetry, occurs via a second-order phase transition. Given that t
The broken symmetry is $Z_2$, corresponding to a transition 
% so the transition may expected to be 
of the conventional Ising type.

We define the order parameter  in terms of $n_A$ and $n_B$, the populations of the $A$ and $B$ sites, as the imbalance in sublattice population:
\be\label{eq:deltan}
\delta n=\frac12\lp \la n_A\ra -\la n_B\ra \rp
,
\ee
where $\la ...\ra$ denotes ensemble average.
We expect that the order parameter $\delta n$ is zero above the transition temperature, $T>T_c$, becoming nonzero at $T<T_c$. The behavior in the vicinity of the transition is governed by the dependence of the free energy $F=E-TS$. Below we analyze the dependence of $F$ vs. $\delta n$, and use it to estimate the ordering temperature $T_c$.
% which we derive and analyze below.

The free energy of a generic state of adatoms is given by a sum of the interaction energy, analyzed in the main text, and the entropic contribution due to adatoms populating the $A$ and $B$ sites,
\be\label{eq:energy_order}
F(n_A,n_B)=E_{\rm int}(n_A,n_B)-\frac{N}2 T s(n_A)-\frac N2T s(n_B),
\ee
%LL where $E_{\rm int}$ is the interaction energy, analyzed in the main text,
where $N$ is the total number of carbon atoms, and $s(n_{A(B)})$ is
 the entropy
%LL density
per site, which is given by
\be\label{eq:entropy}
s(x)=-(x\ln x+(1-x)\ln (1-x)).
\ee
In the limit of low occupancy, which is of interest for us here, this expression is simplified as $s(x)=x\ln e/x$.

To make further progress, we assume that the interaction energy dependence on $\delta n$ can be approximated by a quadratic function,
\be
E_{\rm int}(n_A,n_B)=-\alpha\delta n^2
,\quad
\alpha=\frac{N}{n}\delta E (n)
,
\ee
where $\delta E$ is the energy gain per adatom due to complete sublattice ordering, and  $n=(n_A+n_B)/2$ is the average adatom concentration.
Taylor expanding the entropic part of $F$ in $\delta n$, we find
\bea
&&\frac{N}2\lb (n+\delta n)\ln\frac{e}{n+\delta n}+(n-\delta n)\ln\frac{e}{n-\delta n}\rb \\
&&= Nn\ln\frac{e}{n}-\frac{N}{2n}\delta n^2-\frac{N}{12n^3}\delta n^4+O(\delta n^6)
\eea
Using these results, we can write
%LL Expanding Eq.(\ref{eq:energy_order}) to the fourth order in $\delta n$, we obtain
the effective free energy functional in the vicinity of the phase transition as
\be\label{eq:energy_dis}
\Delta F=F(n+\delta n,n-\delta n)-F(n,n)=N\left[ a\delta n^2+b\delta n^4\right ],
\ee
where
\be\label{eq:a-b}
a=\frac{T}{2n}-\frac{\delta E (n) }{n}, \,\, b=\frac{T}{12 n^3}.
\ee
For the transition temperature, found from the condition $a=0$, this gives
\be\label{eq:Tc}
T_c=2 \delta E (n).
\ee
 The values of $\delta E$ taken from Fig. 2 inset (see the main text) are of the order of several hundred Kelvin. Thus, even for the adatom coverage as small as $n\approx 0.03$, critical temperature values are in the range of hundreds of Kelvin.

\end{document}